\documentclass[prl,twocolumn,showpacs,superscriptaddress]{revtex4}
\usepackage{graphicx,bm,amssymb,amsmath}

\newcommand{\Eq}[1]{Eq.~\eqref{#1}}
\newcommand{\eq}[1]{\eqref{#1}}
\newcommand{\Fig}[1]{Fig.~\ref{#1}}

\newcommand{\beq}{\begin{equation}}
\newcommand{\eeq}{\end{equation}}

\newcommand{\beqa}{\begin{eqnarray}}
\newcommand{\eeqa}{\end{eqnarray}}

\newcommand{\Beqa}{\begin{eqnarray*}}
\newcommand{\Eeqa}{\end{eqnarray*}}
\newcommand{\nn}{\nonumber}

\newcommand{\pdag}{{\phantom{\dagger}}}

\newcommand{\dx}{\partial_x}

\DeclareMathOperator{\sgn}{sgn}

\newcommand{\PRL}[3]{Phys. Rev. Lett.~\textbf{#1}, #2 (#3)}
\newcommand{\PRB}[3]{Phys. Rev. B~\textbf{#1}, #2 (#3)}
\newcommand{\RMP}[3]{Rev. Mod. Phys.~\textbf{#1}, #2 (#3)}
\newcommand{\Science}[3]{Science~\textbf{#1}, #2 (#3)}
\newcommand{\Nature}[3]{Nature~\textbf{#1}, #2 (#3)}
\newcommand{\EPL}[3]{Europhys. Lett.~\textbf{#1}, #2 (#3)}

\newcommand{\JPCM}[3]{J. Phys. Condens. Matter~\textbf{#1}, #2 (#3)}
\newcommand{\SSC}[3]{Sol. State Commun.~\textbf{#1}, #2 (#3)}

\newcommand{\etal}{\textit{et al.}}
\begin{document}

\title{Generation of spin current by Coulomb drag}

\author{M. Pustilnik}
\affiliation{School of Physics, Georgia Institute of Technology,
Atlanta, GA 30332}
\author{E.G. Mishchenko}
\affiliation{Department of Physics, University of Utah,
Salt Lake City, UT 84112}
\author{O.A. Starykh}
\affiliation{Department of Physics, University of Utah,
Salt Lake City, UT 84112}

\begin{abstract}
Coulomb drag between two quantum wires is exponentially
sensitive to the mismatch of their electronic densities. The application
of a magnetic field can compensate this mismatch for electrons of
opposite spin directions in different wires. The resulting enhanced
momentum transfer leads to the conversion of the charge current in the
active wire to the spin current in the passive wire.
\end{abstract}

\maketitle

\begin{figure}[h]
\includegraphics[width=0.44\columnwidth]{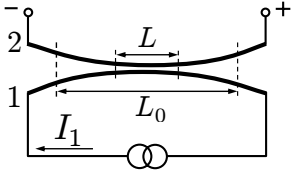}
\caption{
Equivalent circuit for measurement of Coulomb drag between
two quantum wires. Coulomb drag manifests itself in the appearence
of the potential difference $V_2$ between the ends of the open circuit of
which the passive wire $2$ is a part ($V_2$ is positive
if it has the polarity indicated). % in the figure.
\label{Fig1}
}
\end{figure}

A set of unusual transport phenomena in which electron-electron
interactions induce transfer of momentum between distinguishable
systems of fermions is known as Coulomb drag effect. 
Conventional Coulomb drag~\cite{PP} occurs between two spatially
separated conductors. In the standard setup, see
\Fig{Fig1}, dc current $I_1$ flows through the \textit{active}
conductor $1$ inducing a voltage drop $V_2$ in the \textit{passive}
conductor $2$. Quantitatively, the effect is characterized by the 
dimensionless \textit{drag resistance}
\beq
R_d = \lim_{I_1\to \,0}
(e^2\!/h)V_2/I_1.
\label{1}
\eeq
Unlike the usual two-terminal resistance, $R_d$ is sensitive to electronic
correlations within the conductors. Therefore, Coulomb drag effect
provides an important tool to probe these correlations. Coulomb drag
was observed experimentally in two-dimensional bilayers~\cite{2DdragEXP}
and, more recently, in one-dimensional quantum wires~\cite{1DdragEXP}.

%%%%%%%%%%%%%%%%%%%%%%%%%%%%%%%%
%%%%%%%%%%%%%%%%%%%%%%%%%%%%%%%%

A different Coulomb drag-type effect, the \textit{spin drag},
originates in momentum transfer between  spin-up and spin-down
electrons \textit{within the same conductor}~\cite{SpinDrag}.
The spin drag provides a non-dissipative mechanism of relaxation
of a pure spin current.
%(Note that unlike the charge current, the spin current in a clean system
%is not ``protected'' by momentum conservation).
Interactions are therefore \textit{destructive} for spin currents. 
Because robust generation of spin currents is important in view 
of possible applications in spintronics~\cite{zutic}, the limitations 
arising due to the spin drag effect are now a subject of active 
research~\cite{SpinDrag,SpinDragEXP}.

In this paper we demonstrate that interactions can {\it induce} spin
current rather than suppress it. This is possible in a novel type of
Coulomb drag effect, interaction-induced transfer of momentum
between spin-up and spin-down electrons that belong to \textit{separate}
conductors. We show that this effect can be realized in the standard
setting of Coulomb drag between two clean quantum wires in a magnetic
field~\cite{1DdragEXP}.
While the electric current $I_2$ in the passive wire is zero, the spin current
$I_{2s}=I_{2\uparrow} - I_{2\downarrow}$ can flow~\cite{relaxation}, 
i.e. the system acts as a \textit{charge current to spin current converter}. 
The efficiency of the conversion can be characterized by the ratio
\beq
C = I_{2s}/I_1.
\label{2}
\eeq
Below we show that the drag resistance $R_d$ has a
maximum at a certain value $B_0$ of Zeeman energy. For
\beq
\max\{T,|B-B_0|\}\ll B_0
\label{3}
\eeq
the conversion efficiency $C\sim R_d$ [see Eqs. \eq{19} and \eq{24}],
and the dependence of $R_d$ on temperature $T$ is described by a power
law with the exponent depending on the interaction strength, see \Eq{16}.
For sufficiently strong interaction the power-law dependence crosses
over to $R_d\sim 1$ at very low temperatures. We start with a heuristic
explanation of the origin of the effect, and then proceed with the derivation
of the results.

If the electronic densities in the wires $n_1$ and $n_2$ were equal,
the dominant contribution to $R_d$ at low temperatures would come
from  processes with large momentum transfer between the wires
(backscattering), which may result in a finite $R_d$ in the limit
$T\to 0$~\cite{NA,F,KS}. In reality, however, the densities are
always slightly different,
\[
|n_1-n_2|\ll n,
\quad
n= (n_1+n_2)\!/2
\]
(let us assume that $n_1<n_2$), so that the corresponding Fermi
momenta $k_{1,2} = \pi n_{1,2}/2$ are different as well. In this
case, the backscattering contribution to $R_d$ is exponentially
suppressed at low temperatures~\cite{FKS,drag}.

The suppression is easy to understand as follows. To the
lowest order in the strength of the interwire interaction, the
backscattering contribution to $R_d$ can be written as~\cite{drag,ZM}
\beq
\frac{R_d}{L}\sim\frac{U_{2k}^2}{T}
\!\int\!dq\!\int_0^\infty \!\!\!d\omega\,
e^{-\omega/T}\prod_i S_i^{2k}(q,\omega)
\label{4}
\eeq
Here $L$ is the length of the region in which the wires
interact with each other (see \Fig{Fig1}),
$U_{2k}$ is $2k$-Fourier component of the interwire
interaction potential (with $k=(k_1 + k_2)/2 = \pi n/2$), and
$S_i^{2k}(q,\omega)=S_i(q,\omega)\bigr\rvert_{q\sim 2k}$
is the Fourier transform of the dynamic structure factor
$S_i(x,t) = \bigl\langle\rho_i(x,t)\rho_i(0,0)\bigr\rangle$
(here $\rho_i$ is the local density operator for wire $i$).

%%%%%%%%%%%%%%%%%%%%%%%%%%%%%%%%
\begin{figure}[h]
\includegraphics[width=0.9\columnwidth]{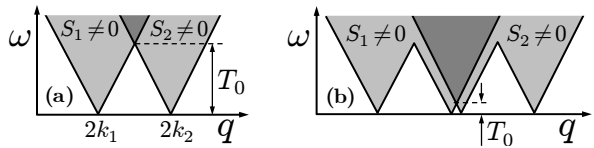}
\caption{
(a) Regions in $(\omega,q)$-plane where $S_{1,2}>0$ at $T=0$ and
$q\sim 2k$. The dark triangle indicates the region where
$S_1 S_2>0$.
(b) In a magnetic field, the low-energy sectors in $S_i(q,\omega)$ split
in two, which leads to the decrease of $T_0$, the minimal energy at which
$S_1$ and $S_2$ overlap at $T=0$.
\label{Fig2}
}
\end{figure}
%%%%%%%%%%%%%%%%%%%%%%%%%%%%%%%%

At $T=0$ and $q\sim 2k$, the two structure factors overlap only at
$\omega>T_0\sim v|k_1-k_2|$, where $v = \pi n/2m$ is the ``average''
Fermi velocity, see \Fig{Fig2}(a). Because of the factor $e^{-\omega/T}$
in \Eq{4}, this translates to the activational temperature
dependence of the drag resistance, $R_d \propto e^{-T_0/T}\!$.
Although at any $T>0$ the structure factors are finite for
all $\omega$ and $q$,  the ``leakage'' of the spectral weight
beyond the boundaries indicated in \Fig{Fig2}(a) affects only the
power-law prefactor in the expression for $R_d$.

With the backscattering contribution exponentially suppressed, 
$R_d$ is dominated by small momentum transfer and vanishes 
at $T\to 0$ as $R_d\propto T^5$~\cite{drag}. In principle, the
densities can be fine-tuned to be equal, which would increase the 
backscattering contribution. Another possibility, which leads 
to spin current generation, is to place the system in a magnetic field.

In a field the single-particle energies $\xi_{k\sigma}$ of the
spin-up $(\uparrow)$ and spin-down $(\downarrow)$ electrons
(labeled by $\sigma = \pm 1$) include Zeeman contribution
$\delta\xi_{k\sigma}=\sigma B/2$.
As a result, $n_{i\downarrow}> n_{i\uparrow}$, and the Fermi
momenta are
\beq
k_{i\sigma} = k_i - \sigma \delta k/2
\label{5}
\eeq
with $\delta k(B)\sim B/v$ (see below). For each wire, the low-energy
sector in $S_i^{2k}(q,\omega)$ then splits in two, located at $q =2k_{i\sigma}$,
see \Fig{Fig2}(b). The scale $T_0$ is $B$-dependent and vanishes
at a certain field $B_0$,  $T_0(B)\sim |B-B_0|$ (see \Eq{12} below).
At $|B-B_0|\lesssim T$, the backscattering contribution to $R_d$
is no longer exponentially suppressed and dominates at sufficiently 
low temperatures. Moreover, in the regime \eq{3} the main contribution 
to the integral in \Eq{4} comes from the overlap of $S_{1\downarrow}$ 
and $S_{2\uparrow}$. In other words, \textit{almost all of the momentum 
is transferred from spin-down electrons in the active wire to spin-up 
electrons in the passive one}. Therefore, both $R_d$ and $C$ will 
have a maximum at $B=B_0$.

We evaluate $R_d$ and $C$ in the regime \eq{3} using the
bosonization technique ~\cite{1D_books}. At energies well below
$B_0$, which in turn is small compared with the Fermi energy
$\epsilon_F$, the wire $i$ $(i=1,2)$ is described by the Hamiltonian
\beq
H_i = \sum_{m}\frac{v_m}{2}\!\int\!dx \!
\left[
g_m^{-1}(\dx\varphi_{im})^2 + g_m (\dx\vartheta_{im})^2
\right],
\label{6}
\eeq
where $m=c,s$ labels the charge (spin) modes, and the
bosonic fields satisfy
\beq
\bigl[\varphi_{im}(x),\vartheta_{i'm'}(y)\bigr]
= (i/2)\delta_{ii'}\delta_{mm'}\sgn(x-y).
\label{7}
\eeq
For simplicity, we assume that both wires are described by the
same set of parameters $\{v_m, g_m\}$. These parameters are
related to each other according to
\beq
g_c = v/v_c,
\quad
g_s(B_0) = 1 + [2\ln(\epsilon_F\!/B_0)]^{-1}
\label{8}
\eeq
(so that $1-g_c \gg g_s-1>0$ for $B_0\ll\epsilon_F$), and the
velocities $v_c>v$ and $v_s<v$ can be further expressed
in terms of the interaction within the wires~\cite{1D_books}.

Fermion operators in the bosonic representation are
\beq
\psi_{i\alpha\sigma}(x) = \mu_{i\alpha\sigma}\sqrt{p_0}\,
e^{i\alpha[\eta_{i\alpha\sigma}(x) + k_{i\sigma} x]}.
\label{10}
\eeq
Here $\alpha = +1(-1)$ for the right (left) moving fermions,
$\mu_{i\alpha\sigma}^\pdag=\mu_{i\alpha\sigma}^\dagger$ are real
(Majorana) fermions that satisfy
$\bigl\lbrace\mu_{i\alpha\sigma},\mu_{i'\alpha'\sigma'}\bigr\rbrace
= 2\delta_{ii'}\delta_{\alpha\alpha'}\delta_{\sigma\sigma'}$
(these operators enforce correct anticommutation relations between
different fermionic species), $p_0\sim B_0/v$ is the high-momentum
cutoff, and $\eta_{i\alpha\sigma}$ is a linear combination of
$\varphi_{im},\vartheta_{im}$, which in the leading order in
$B_0/\epsilon_F\ll 1$ is given by~\cite{field}
\beq
\eta_{i\alpha\sigma}
= \sqrt{\pi/2} \,\bigl(\varphi_{ic} + \alpha\vartheta_{ic}
+ \sigma\varphi_{is}+\alpha\sigma\vartheta_{is}\bigr).
\label{11}
\eeq
Fermi momenta $k_{i\sigma}$ in \Eq{10} are given  by \Eq{5}
with $ \delta k(B) = g_s B/v_s$, and $T_0(B)$ (see \Fig{Fig2})
at $B\to B_0$ is
\beq
T_0(B) \approx g_s|B-B_0|,
\quad B_0 \approx v_s|k_2-k_1|
\label{12}
\eeq
($B_0$ is the root of the equation $g_s(B) B = v_s|k_2-k_1|$).

With the help of \Eq{10}, the $2k$-harmonic of the density
operator $\rho_i^{2k} = \sum_\sigma\!\rho_{i\sigma}^{2k}$
is written as
\[
\rho_{i\sigma}^{2k} = p_0 \mu_{i\sigma}
\exp\bigl[i\sqrt{2\pi}\,(\varphi_{ic}+\sigma\varphi_{is}) + 2ik_{i\sigma}x\bigr]
+ \text{H.c.},
\]
where $\mu_{i\sigma} = \mu_{i,-1,\sigma}\mu_{i,+1,\sigma}$.
Since the Hamiltonian \eq{6} is quadratic, evaluation of the structure
factor is straightforward~\cite{1D_books} and yields
$S_i(x,t) = \sum_{\sigma} S_{i\sigma}(x,t)$ with
\[
S_{i\sigma}(x,t) = 2p_0^2 \cos(2k_{i\sigma}x)\! \prod_{\alpha,m}
\left[ \frac{T/(2 p_0 v_m)}{\sinh\bigl(\pi T
\tau_{\alpha m}\bigr)} \right] ^{g_m/2},
\]
where $\tau_{\alpha m} = x/v_m - \alpha (t-i0)$.

As discussed above, the condition \eq{3} ensures that the main
contribution to the integral in \Eq{4} comes from the nonvanishing
overlap of $S_{1\downarrow}$ and $S_{2\uparrow}$; the remaining
contributions are suppressed as $\propto\exp(-B_0/T)$. In order to
evaluate $R_d$, it is convenient to convert \Eq{4} to
space-time representation,
\beq
R_d/L = (\pi/2)U_{2k}^2\!\int_{-\infty}^\infty\!dx\,dt\,(it)\,
S_1(x,t)\, S_2(x,t).
\label{13}
\eeq
Substituting here $S_{1\downarrow}$ for $S_1$ and and $S_{2\uparrow}$
for $S_2$, we find
\beq
R_d\sim n\lambda^2_{2k}L
\,\frac{B_0}{\epsilon_F} \left[\frac{|B-B_0|}{B_0}\right]^{4g-3}
\!\! F\! \left(\frac{g_s|B-B_0|}{T}\right),
\label{14}
\eeq
where $\lambda_{2k} = U_{2k}/2\pi v$ and $ g=(g_c+g_s)\!/2$.
The function $F(z)$ in \Eq{14} is given by
\beqa
F(z) &=& \iint\!
\frac{(z/2)^{3-4g}\exp(2iz\xi/\pi)\,d\xi d\zeta}
{
\prod_m\bigl[\cosh\bigl(\frac{v_s}{v_m}\,\xi +\zeta\bigr)
\cosh\bigl(\frac{v_s}{v_m}\,\xi -\zeta\bigr)\bigr]^{g_m}
}
\nn\\
\nn\\
&&\quad\sim\, \left\lbrace
\begin{array}{lc}
z^{3-4g}, & z\ll 1 \\
z e^{-z}, & 1\ll z\ll z_0 \\
z^{1-2g_c} e^{-z}, & z\gg z_0
\end{array}
\right., \label{15} \eeqa where $z_0 =
g_c(\pi/2)(v_s/v_c)\tan\bigl[(\pi/2)(v_s/v_c)\bigr]$ (so that
$z_0\sim (1-g_c)^{-1}\gg 1$ for weak interaction). In deriving Eqs.
\eq{14},\eq{15} we changed the integration variables in \eq{13} to
$\xi = \pi T x/v_s$ and $\zeta = \pi T t$, shifted the path of
integration over $\zeta$ off the real axis by $-i\pi/2$, and
evaluated the resulting integral in the saddle-point approximation.
According to Eqs. \eq{14},\eq{15}, and in agreement with the
discussion above, $R_d(B)$ has a narrow peak of the width $\delta
B\sim T\ll B_0$ at $B=B_0$. Its height is given by
\beq
\max\bigl\lbrace R_d(B)\bigr\rbrace\sim 
n\lambda^2_{2k}L\, (B_0/\epsilon_F)(T\!/B_0)^{4g-3}.
\label{16}
\eeq

Note that the difference between $v_s$ and $v_c$ is important 
only at large $|B-B_0| \gtrsim T/(1-g_c)$. In the opposite limit 
one can set $v_s/v_c\to 1$, which yields 
$F(z) = \left|\Gamma\!\left(g+iz\!/2\pi\right)\right|^4\!/\Gamma^2(2g)$,
in agreement with \Eq{15}; the corresponding $T$-dependence
is exactly the same as that for the drag between two spinless 
wires~\cite{FKS,Fiete}.

In order to relate the conversion efficiency \eq{2} to the drag
resistance \eq{16}, we note that as far as the passive wire is
concerned, in the regime~\eq{3} Coulomb drag induces the electric
field that couples to spin-up electrons only. The effect of this
field can be described by adding to the Hamiltonian of the passive
wire a term
\beq
\delta H_2 = e\!\int\!dx\,\Phi_d(x)
\rho_{2\uparrow}(x)
= e\!\int\!dx\,\frac{\Phi_d}{2}\,
(\rho_{2c} +\rho_{2s}),
\label{17}
\eeq
where $\Phi_d(x)$ is drag-induced potential, and $\rho_{2c}$ and
$\rho_{2s}$ are charge and spin densities. The potential $\Phi_d(x)$
changes within the region of the length $L$ in which the wires interact
with each other. Assuming that the wires are long, $L_0\gg L$, the
charge and spin currents in response to $\delta H_2$ can be written
as~\cite{KF}
\beq
I_{2c} = (2e^2\!/h)\,g_c \,\delta\Phi_d/2,
\quad
I_{2s} = (2e^2\!/h)\,g_s\,\delta\Phi_d/2,
\label{18}
\eeq
where $\delta\Phi_d = \Phi_d(-\infty) - \Phi_d(\infty)$. In writing
\Eq{18} we took into account the renormalization of the corresponding
conductances by interactions within the wire~\cite{KF}.

On the other hand, the electrostatic potential difference $V_2$
induces charge current $I_V = (2e^2\!/h)V_2$. Here we assumed
that the interactions are efficiently screened within the leads and that
the contacts between the leads and the wires are reflectionless; the
corresponding conductance is not affected by the interactions~\cite{M}.
The condition of vanishing of the total electric current,
$I_2 = I_V + I_{2c} = 0$, then yields $\delta\Phi_d = -2V_2/g_c$.
Eqs.~\eq{1},\eq{2} and \eq{18} now give
\beq
C = I_{2s}/I_1 = 2(g_s/g_c)R_d.
\label{19}
\eeq
Thus, under the conditions \eq{3} the dependence of conversion
efficiency $C$ on $B$ and $T$ is indeed the same as that of the drag resistance
$R_d$, as asserted above.

\Eq{19} does not account for the reduction of $I_s$ due to the
momentum transfer between the two spin subsystems within the
passive wire (spin drag). Indeed, in the framework of the
Tomonaga-Luttinger model \eq{6} the only source of spin drag is
the backscattering in the spin sector, which at $T\ll B$ is exponentially
suppressed. The dominant contribution to spin drag then comes from
the processes with small momentum transfer. Accounting for these
processes requires explicit consideration of the nonlinearity of the
electronic spectrum~\cite{drag}. Proceeding along the lines of~\cite{drag},
we found the corresponding correction to the spin current $I_{2s}$
at $T\ll B$ and in the lowest non-vanishing order in the interaction
strength,
\beq
\delta I_{2s}/I_{2s}
\sim - nL_0(1-g_c)^4 (B/\epsilon_F)^4(T/B)^5.
\label{20}
\eeq
In writing \Eq{20} we took into account that Fermi velocities for
spin-up and spin-down electrons differ by $\delta v\sim B/k\ll v$.
The correction \eq{20} is small and does not affect the validity
of \Eq{19}.

The above consideration is based on the perturbative expression \Eq{4}.
In order to analyze the relevance of the higher-order contributions,
we introduce new fields
\[
\phi_c = 2^{-1/2}(\varphi_{1c}-\varphi_{2c}),
\quad
\phi_s = 2^{-1/2}(\varphi_{1s}+\varphi_{2s}),
\]
and similarly defined $\theta_c$ and $\theta_s$. The fields
obey the commutation relations analogous to
%$\bigl[\phi_{m}(x),\theta_{m}(y)\bigr] = (i/2)\sgn(x-y)$, cf.
\Eq{7}, and their dynamics is governed by the Hamiltonian
$H = \int\!dx\,\mathcal{H}$ with
\beqa
\mathcal{H} \!&=&\!\sum_{m}\!\frac{v_m}{2}
%\!\int\!dx\!
\left[ g_m^{-1}(\dx\phi_{m})^2 + g_m (\dx\theta_{m})^2
\right]-\,2v\lambda_0
%\!\int_0^L\!\!dx\,
(\dx\phi_c)^2 \label{25}
\nonumber\\
&&
%\!\int_0^L\!\!dx
+\,\,4\pi v \lambda_{2k} p_0^2
\cos\Bigl\{\!\sqrt{4\pi}\,(\phi_c -\phi_s) + 2K_0 x\Bigr\}.
\label{21}
\eeqa
The second and the third terms here describe, respectively, the
forward and backward scattering between the spin-up electrons
in wire $2$ and the spin-down electrons in wire $1$, with
$\lambda_0$ defined similarly to $\lambda_{2k}$ in \Eq{14}, and
$K_0 = T_0(B)/v_s$.

The forward scattering term in \Eq{25} leads to small corrections
to $v_c$ and $g_c$,
$
\delta g_c/g_c \approx -\delta v_c/v_c
\approx 2g_c^2 \lambda_0 \ll 1,
$
which modify the exponent in Eqs. \eq{14}-\eq{16}, %~
$g\to g+\delta g_c/2$. %, see \Eq{16}.
The backscattering, however, can be relevant in the renormalization
group sense~\cite{RG}. For $L\to\infty$ and $K_0\to 0$ it then
results in the opening of a gap
\beq
\Delta\!\sim\!B_0\lambda_{2k}^{1/(2-2g)}
\label{22}
\eeq
in the excitation spectrum. The gapped state is the
``zigzag''-ordered state formed by the spin-down electrons in
wire $1$ and the spin-up electrons in wire $2$.

The gap remains open for finite $K_0$ as long as the energy gained
due to its formation is sufficient to overcome the cost of the adjustment
of the densities needed to form the zigzag order. In the context of
quantum wires such adjustment (known as commensurate-incommensurate
transition) was discussed recently in~\cite{SMHG,FKS}. The adjustment
takes place at not too large $K_0$, $K_0<K_c\sim \Delta/v$, and occurs
even when $L$ is finite. As a result, the width $\delta B$ of the peak in
$R_d(B)$ saturates at low temperatures,
\beq
\delta B\sim \max\{T, \Delta\}.
\label{23}
\eeq

For $L\ll v/\Delta$ the zigzag order can not be formed and \Eq{16}
is applicable. In this case $\max\bigl\lbrace R_d(B)\bigr\rbrace\ll 1$
for all $T$. The higher-order contributions become important for 
$L\gtrsim v/\Delta$ and at $T\lesssim\Delta$~\cite{NA,F,KS,FKS}. 
While finding the detailed dependence $R_d(T)$ in this regime is 
beyond the scope of this Letter, the limiting values of $R_d$ and 
$C$ at $T\to 0$ can be found as follows. 
Imagine that the two wires are connected to noninteracting
reservoirs and  a bias is applied only to the electrons with spin
$\sigma$ in wire $i$. The resulting current of electrons with spin
$\sigma'$ in wire $i'$ is
$I_{i'\!\sigma'} = G_{i'\!\sigma'\!,i\sigma}V_{i\sigma}$,
where
$G_{i'\!\sigma'\!,i\sigma} = G_{i\sigma,i'\!\sigma'}$
is the corresponding conductance. At $T\to 0$ the spin-up
electrons in wire $2$ are ``locked'' with the spin-down electrons
in wire $1$, and we expect that
$G_{1\downarrow,1\downarrow},G_{2\uparrow,2\uparrow},
G_{1\downarrow,2\uparrow}\to e^2\!/2h$. At the same time,
$G_{1\uparrow,1\uparrow},G_{2\downarrow,2\downarrow} \to e^2\!/h$,
while $G_{1\uparrow,2\downarrow}\to 0$. Setting
$V_{i\sigma} = V_i$, $I_i = I_{i\uparrow} + I_{i\downarrow}$,
%and using Eqs. \eq{1},\eq{2},
we find
\beq
R_d\to 1/4,
\quad
C\to 1/2.
\label{24}
\eeq

To conclude, we showed that in the presence of the applied magnetic
field the standard Coulomb drag measurement setup acts as a charge
current to spin current converter. Both the drag resistance and the
conversion efficiency exhibit a maximum at a certain value of the
field controlled by the density mismatch between the wires.

Our results are applicable for long $(kL_0\gg 1)$ ballistic quantum
wires. The wires studied in~\cite{1DdragEXP} exhibit a well-defined
conductance quantization, which guarantees that the elastic mean free
path exceeds the length of the wires $L_0$.
While it is very plausible that $kL_0\gg 1$ for at least some of
the samples studied in~\cite{1DdragEXP} (with $L_0$ ranging from
$0.4\,\mu\text{m}$ to $4\,\mu\text{m}$), the density of
electrons in these wires is difficult to estimate. Fortunately, such estimate
is available for the coupled-wire system studied in~\cite{Amir}:
$L\approx L_0\approx 10\,\mu\text{m}$ and $kL_0\sim 10^3$. 
Although the experiments~\cite{Amir} focus on the
momentum-resolved tunneling, the same system can be
employed to study the Coulomb drag effect as well.
For this system, the typical density mismatch $|n_1-n_2|/n\sim 10^{-2}$
corresponds to $B_0\sim 1\,\text{K}$ (which amounts to the
applied field of $\sim 3\,\text{Tesla}$), hence
the regime \eq{3} is well within the reach of the experiments.

\begin{acknowledgments}
We thank L. Glazman and G. Vignale for useful discussions. MP and
EGM are grateful to the Kavli Institute for Theoretical Physics at
UCSB and MP thanks the Aspen Center for Physics for their
hospitality. This work is supported by the NSF (grants DMR-0503172
and DMR-0604107), by the DOE (grants DE-FG02-ER46311 and
DE-FG02-06ER46313), and by the ACS PRF (grant 43219-AC10).
\end{acknowledgments}


\begin{thebibliography}{99}
\bibitem{relaxation}
A spin current $I_{s}$ in a wire leads to the spin accumulation
$M = N_\uparrow-N_\downarrow\neq 0$ in the reservoirs to which 
the wire is attached ($N_\sigma$ is number of electrons with spin 
$\sigma$ in the reservoir). In the steady state $M = I_s\tau/e$, where 
$1/\tau$ is the spin relaxation rate in the reservoirs.
\bibitem{RG}
By power counting, the backscattering in \Eq{21} is relevant
if $g<1$. For weak interaction and large interwire separation
$\lambda_0\ll 1-g_c\ll 1$, and $g<1$ at any $B\ll
\epsilon_F$. For small interwire separation
the situation is more delicate and requires the analysis of the RG flow
similar to that in Ref.~\cite{KS} at $B=0$.
\end{thebibliography}
\end{document}